\documentstyle[12pt]{article}
\def\journal{\topmargin .3in    \oddsidemargin .5in
        \headheight 0pt \headsep 0pt
        \textwidth 5.625in 
        \textheight 8.25in 
        \marginparwidth 1.5in
        \parindent 2em
        \parskip .5ex plus .1ex         \jot = 1.5ex}
\journal

\catcode`\@=11

\catcode`\@=11
\def\section{\@startsection {section}{1}{0pt}{-3.5ex plus -1ex minus
 -.2ex}{2.3ex plus .2ex}{\raggedright\large\bf}}
\catcode`\@=12
\def\e{\varepsilon}
\def\ie{\hbox{\it i.e.\/}}

\def\bra #1{\left\langle #1\right|}
\def\ket #1{\left| #1\right\rangle}
\def\vev #1{\left\langle #1\right\rangle}
\def\abs#1{\left| #1\right|}
\newtheorem{fh}{Theorem}
\newtheorem{gvth}[fh]{Theorem}
\begin{document}
\begin{titlepage}
\begin{center}
September 11, 1992     \hfill    LBL-32872 \\
                       \hfill    UCB-PTH-92/33 \\

\vskip .5in

{\large \bf Meson Mass Splittings in the Nonrelativistic Model}
\footnote{This work was supported by the Director, Office of Energy
Research, Office of High Energy and Nuclear Physics, Division of High
Energy Physics of the U.S. Department of Energy under Contract
DE-AC03-76SF00098.}

\vskip .5in

Richard F. Lebed\\[.5in]

{\em  Department of Physics\\
      University of California\\
      and\\
      Theoretical Physics Group\\
      Physics Division\\
      Lawrence Berkeley Laboratory\\
      1 Cyclotron Road\\
      Berkeley, California 94720}
\end{center}

\vskip .5in

\begin{abstract}
Mass splittings between isodoublet meson pairs and between $0^{-}$ and
$1^{-}$ mesons of the same valence quark content are computed in a detailed
nonrelativistic model. The field theoretic expressions for such splittings are
shown to reduce to kinematic and Breit-Fermi terms in the nonrelativistic
limit.
Algebraic results thus obtained are applied to the specific case of the
linear-plus-Coulomb potential, with resultant numbers compared to experiment.

\end{abstract}
\end{titlepage}
\renewcommand{\thepage}{\roman{page}}
\setcounter{page}{2}
\mbox{ }

\vskip 1in

\begin{center}
{\bf Disclaimer}
\end{center}

\vskip .2in

\begin{scriptsize}
\begin{quotation}
This document was prepared as an account of work sponsored by the United
States Government.  Neither the United States Government nor any agency
thereof, nor The Regents of the University of California, nor any of their
employees, makes any warranty, express or implied, or assumes any legal
liability or responsibility for the accuracy, completeness, or usefulness
of any information, apparatus, product, or process disclosed, or represents
that its use would not infringe privately owned rights.  Reference herein
to any specific commercial products process, or service by its trade name,
trademark, manufacturer, or otherwise, does not necessarily constitute or
imply its endorsement, recommendation, or favoring by the United States
Government or any agency thereof, or The Regents of the University of
California.  The views and opinions of authors expressed herein do not
necessarily state or reflect those of the United States Government or any
agency thereof of The Regents of the University of California and shall
not be used for advertising or product endorsement purposes.
\end{quotation}
\end{scriptsize}

\vskip 2in

\begin{center}
\begin{small}
{\it Lawrence Berkeley Laboratory is an equal opportunity employer.}
\end{small}
\end{center}

\newpage
\renewcommand{\thepage}{\arabic{page}}
\setcounter{page}{1}
\section{Introduction}
{\indent}The splitting of the masses of mesons in an isospin doublet, sometimes
called electromagnetic splitting, has traditionally been attributed primarily
to explicit isospin breaking ({\ie} $m_{u} \neq m_{d}$) and differences between
the charges of the valence quark-antiquark pairs ($Q_{u} \neq Q_{d}$),
with hyperfine, spin-orbit, and other effects neglected in
comparison.  Such a model serves to explain the observed splittings $K^{0} -
K^{+} = 4.024 \pm 0.032$ MeV and $D^{+} - D^{0} = 4.77 \pm 0.27$ MeV, but has
failed in light of the surprisingly small $B^{0} - B^{+} = 0.1 \pm 0.8$ MeV.

It is precisely this mass difference which has led to the proposal of a variety
of models.  Some of these \cite{Singh,Chan,Kim,Flamm} are based on the
nonrelativistic model of hadron masses put forth by De R\'{u}jula, Georgi, and
Glashow \cite{DeRujula} soon after the development of QCD.  Such models have
the
unfortunate tendency to predict numbers no smaller than
$B^{0} - B^{+} \simeq 2$ MeV, well outside the current experimental limits.
Using more phenomenological models \cite{Tiwari,Goity}, one can obtain a
smaller
splitting in closer agreement with experiment.  Nevertheless, it may seem odd
that the usual nonrelativistic model, which works well for the {\em D\/}- and
even the {\em K\/}- mesons, should fail in the case of the {\em B\/}, which
boasts an even heavier quark.

The primary conclusion of this work is that it {\em is\/} possible to explain
the mass splittings of heavy mesons ($D$ and $B$, but not $K$) in an ordinary
nonrelativistic model, as long as we take into account {\em all\/} corrections
to consistent orders of magnitude, that expectation values of the mesonic
wavefunctions in general have mass dependence, and that the running of the
strong coupling constant is not negligible.

In this spirit, the paper is organized as follows: in the second section we
consider the problem of computing mesonic mass contributions in field theory.
Then, in the third section, we demonstrate that the nonrelativistic limit of
the field-theoretic result leads to kinematic terms and the Breit-Fermi
interaction, exactly as stated in De R\'{u}jula {\it et al}.  This is followed
in Section 4 by an exhibition of the full mass splitting relations for
isodoublet $0^{-}$ and $1^{-}$ meson pairs, as well as ($0^{-}$, $1^{-}$) pairs
with the same valence quarks.  Section 5 discusses the application of
quantum-mechanical theorems, including a very useful generalized virial
theorem,
to the problem of reducing the number of independent expectation values in the
splitting formulae.  These theorems are applied to the popular choice of a
linear-plus-Coulomb potential in Section 6, with numerical results presented in
Section 7.

\section{Mass Computation in Field Theory}
{\indent}Typically, the computation of mesonic mass splittings in a
nonrelativistic model is accomplished by starting with the Breit-Fermi
interaction~\cite[Secs. 38-42]{Bethe}
\begin{eqnarray}
\lefteqn{H_{BF} = \label{bf}
\sum_{i>j} (\alpha Q_{i} Q_{j} + k \alpha_{s} )} & & \nonumber \\
& & \left\{
\frac{1}{\abs{\vec{r}_{ij}}} -
\frac{1}{2m_{i}m_{j} \abs{\vec{r}_{ij}}}
( \vec{p}_{i} \cdot \vec{p}_{j} +
\hat{r}_{ij} \cdot (\hat{r}_{ij} \cdot \vec{p}_{i} ) \vec{p}_{j} )
\right. \nonumber \\
& & - \frac{\pi}{2} \delta^3(\vec{r}_{ij})
\left[ \frac{1}{m_{i}^{2}} + \frac{1}{m_{j}^{2}} +
\frac{4}{m_{i} m_{j}} \left( \frac{4}{3} \vec{s}_{i} \cdot \vec{s}_{j}
+ \left( \frac{3}{4} +\vec{s}_{i} \cdot \vec{s}_{j} \right)
\delta_{q_{i} \bar{q_{j}}} \right) \right]
\nonumber \\
& & - \frac{1}{2 \abs{\vec{r}_{ij}}^{3}} \left[
\frac{1}{m_{i}^{2}} \vec{r}_{ij} \times \vec{p}_{i} \cdot \vec{s}_{i} -
\frac{1}{m_{j}^{2}} \vec{r}_{ij} \times \vec{p}_{j} \cdot \vec{s}_{j}
\right. \nonumber \\
& & \left. \left. + \frac{2}{m_{i} m_{j}} \left(
\vec{r}_{ij} \times \vec{p}_{i} \cdot \vec{s}_{j} -
\vec{r}_{ij} \times \vec{p}_{j} \cdot \vec{s}_{i} +
3(\vec{s}_{i} \cdot \hat{r}_{ij}) (\vec{s}_{j} \cdot \hat{r}_{ij}) -
\vec{s}_{i} \cdot \vec{s}_{j}
\right) \right] \right\},
\end{eqnarray}
where $\vec{r}_{i}$, $\vec{p}_{i}$, $m_{i}$, $\vec{s}_{i}$, and $Q_{i}$ denote
the coordinate, momentum, (constituent) mass, spin, and charge (in units of the
protonic charge) of the {\em i\/}th
quark, respectively; $\vec{r}_{ij} \equiv \vec{r}_{i} - \vec{r}_{j}$;
$\alpha$ and $\alpha_{s}$ are the (running) QED and QCD coupling constants; and
$k=-\frac{4}{3}$ $(-\frac{2}{3})$
is a color binding factor for mesons (baryons).  This expression includes an
annihilation term if $q_{i} = \bar{q_{j}}$ are in a relative $j=1$ state.
{}From this, one chooses
the terms that are considered significant and then calculates the appropriate
quantum mechanical expectation values.  We will pursue this course of
action in the next section; however, this author feels that it would be
worthwhile to first consider the derivation of this interaction for the mesonic
system from the more fundamental field theories of QED and QCD, since this
approach entails greater generality and may provide impetus for work beyond the
scope of this paper.

    We first consider the question of the mass of a composite system from the
point of view of the {\em S}-matrix and interaction-picture perturbation
theory.
The mass of a system, defined as the expectation value of the total Hamiltonian
in the center-of-momentum frame of the constituents, receives contributions
from both the noninteracting and interacting pieces of the Hamiltonian; the
former gives rise to the masses and kinetic energies of the constituents, and
the latter produces the interaction energy.  Technically, the matrix element of
the noninteracting piece {\it in the interaction picture\/} produces terms
which contribute to interactions between {\it renormalized} constituents.
Thus one may think of interactions between ``dressed'' constituents, a topic to
which we return momentarily.

Let us follow the method of Gupta~\cite{Gupta} to derive the interaction
potential from the field-theoretical interaction Hamiltonian.  We begin by
writing the $S$-matrix in the Cayley form
\begin{equation}
S = \frac{1 - \frac{1}{2} iK}{1 + \frac{1}{2} iK} ,
\end{equation}
and expand the Hermitian operator $K$:
\begin{equation}
K = \sum_{n} K_{n} .
\end{equation}
The purpose of this expansion, rather than expanding $S\/$ directly, is to
preserve unitarity in each partial sum of $S$.  The physical effect of this
parametrization is to eliminate diagrams with real intermediate states from the
$S$-matrix expansion.

Computing the terms $K_{n}$, one finds
\begin{equation}
K_{1} = \int_{-\infty}^{+\infty} dt \, H^{I}_{int} (t) ,
\end{equation}
where $I$ indicates the interaction picture.  Now observe that we may invent an
effective Hamiltonian, $H^{I}_{ef \! f}$, such that its first-order
contribution
is equivalent to the contribution from $H^{I}_{int}$ {\it to all orders}.
Thus,
\begin{equation}
K = \int_{-\infty}^{+\infty} dt \, H^{I}_{ef \! f} (t) .
\end{equation}
The interaction energy is then
\begin{equation}
\Delta E = \frac{\bra{f^{I}} H^{I}_{ef \! f} (0) \ket{i^{I}}}{\langle f^{I}
\left. \! \right| i^{I} \rangle} ,
\end{equation}
with $\ket{i^{I}}$ and $\ket{f^{I}}$ actually the same state since the system
is
stable.

In our case, in which $H^{I}_{ef \! f}$ is composed of the interaction terms of
QED and QCD, the lowest-order contribution is $K_{2}$, corresponding to two
interaction vertices: the exchange of one vector boson.  It is easily shown
that
\begin{equation}
K_{2} = iS_{2} = (2 \pi)^{4} \delta^{4} (P_{f} - P_{i}) {\cal M}_{fi}^{(2)} =
\bra{f^{I}} \int dt \, H^{I \, (2)}_{ef \! f} (t) \ket{i^{I}} ,
\end{equation}
where the superscript (2) indicates second-order in the coupling constant, and
$\cal M$ is the usual invariant amplitude for the process.  Eliminating the
delta functions that arise in the right-most expressions we find
\begin{equation}
\Delta E^{(2)} = \frac{\bra{f^{I}} H^{I \, (2)}_{ef \! f} (0) \ket{i^{I}}}
{\langle f^{I} \left. \! \right| i^{I} \rangle} = {\cal M}_{fi}^{(2)} .
\end{equation}
Beyond second-order the relation between interaction energy and the invariant
amplitude becomes less trivial, but nevertheless Gupta has shown that it can be
done.  However, we do not continue to fourth-order in this work, and henceforth
suppress the (2) in the following.

In general, ${\cal M}_{fi}$ at any given order is represented by diagrams of
the form indicated in Figure 1.  The composite state is formed by superposition
of the constituent particle wavefunctions in such a way that the desired
overall quantum numbers for the composite state are obtained.
For the mesonic system, ${\cal M}_{fi}$ is represented by the diagram in Figure
2, where the lowest-order interaction is the exchange of a single gauge boson.
This class of diagrams allows for only the valence quark and antiquark (no sea
$q \overline{q}\/$ pairs or glue), and thus would be a poor model if we chose
these to be current quarks.  Instead, the quarks in our diagrams will be
constituent quarks, and the gauge couplings will assume their running values.
In this way we can model the hadronic cloud, as well as
renormalizations of the lines and vertices of our diagram, so that its
particles are ``dressed'' in two senses.  There is also an annihilation diagram
if the quark and antiquark are of the same flavor.  In this work we
consider only the exchange diagram, since the mesons of greatest interest to us
are those with one heavy and one light quark.

The next step is to obtain the amplitude ${\cal M}_{fi}$, in which the
constituent legs are bound in the composite system, from the Feynman
amplitude $\cal M$ [Figure 3] for the same interaction with {\em free\/}
external constituent legs.  To do this, we need only constrain the free
external legs in a way which reflects the wavefunction and rotational
properties of the meson state.  In general, if the variables $z_{n}\/$ are
the degrees of freedom of the meson state $\ket{\Phi}$, then we may write
\begin{equation}
\ket{\Phi} = \int \hspace{-2.7ex} \sum dz_{n} \, \phi (z_{n}) \, {\cal{O}}
(z_{n}) \ket{0}.
\end{equation}
The function $\phi$ is an amplitude in the variables $z_{n}\/$, {\ie\/} a
wavefunction; and $\cal{O}$ is a collection of Fock space operators which
specifies the rotational properties of $\ket{\Phi}$. The integral-sum symbol
indicates summation over both continuous and discrete $z_{n}$.  In this
notation, we obtain the result
\begin{equation}
\Delta E = \int \hspace{-2.7ex} \sum dz_{f} \, dz_{i} \, \phi^{*} (z_{f})
\phi (z_{i}) f(z_{i} , z_{f}) {\cal{M}} (z_{i} , z_{f}),
\end{equation}
where $f(z_{i},z_{f}) \equiv \bra{0} {\cal{O}}^{\dagger} (z_{f}) {\cal{O}}
(z_{i}) \ket{0}$ is a constraint function.  We have written the energy
contribution in this very general way in order to demonstrate the power of the
technique.

Now we apply this prescription to the usual case of Feynman rules.  Then
$z_{n}$
are quark momenta, $\phi$ is the mesonic momentum-space wavefunction, and $f$
specifies the spin of the meson, as we shall see below.  The energy
contribution
is evaluated in the quark center of momentum frame ({\ie\/} the meson rest
frame), in which the relative momenta of the quark-antiquark pair, intitially
and finally, are denoted by $\vec{p}$ and $\vec{p} \, '$, respectively.
Fourier
transformation of the wavefunctions from momentum-space to position-space
yields
\begin{eqnarray}
\Delta E_{CM} & = & \int \! d^{3} \vec{x}_{f} \int \! d^{3} \vec{x}_{i} \,
\psi^{*}
(\vec{x}_{f}) K(\vec{x}_{f} , \vec{x}_{i}) \psi (\vec{x}_{i}) , \nonumber \\
\lefteqn{\rm where} & &  \nonumber \\
K(\vec{x}_{f} , \vec{x}_{i}) & = & \int \! d^{3} \vec{p} \, ' \int \! d^{3}
\vec{p} \;
{\rm exp} [i(\vec{p} \,' \cdot \vec{x}_{f} - \vec{p} \cdot \vec{x}_{i})] \,
\sum_{spins} f(spins)
{\cal{M}} (\vec{p} \, ' , \vec{p} , spins) , \nonumber
\\
\lefteqn{\rm and} & & \int \! d^{3} \vec{x} \, \psi^{*} (\vec{x}) \psi
(\vec{x})
= 1. \label{deltae}
\end{eqnarray}

As a technical point of fact, it is necessary to keep track of the
normalization
conventions used for wavefunctions, Fourier transforms, and Feynman rules in
order to obtain the true convention-independent $\Delta E\/$.  As it stands,
Eq.~\ref{deltae} locks us into a particular set of Feynman rule normalizations,
which should be made clear in the following expression.  The kinematic
conventions are established in Figure 4.  Then the Feynman amplitude for free
external quark legs and a virtual photon is
\begin{eqnarray}
\lefteqn{
{\cal M} = i \left[ \frac{1}{(2 \pi)^{3/2}} \right] ^{4} \sqrt{\frac{M}{E_{f}}}
\sqrt{\frac{M}{E_{i}}} \sqrt{\frac{m}{\e_{f}}} \sqrt{\frac{m}{\e_{i}}} }
& & \nonumber \\
& & \left[ \bar{v}_{{H}_{i}} (\vec{P}_{i}) (-iQe \gamma_{\mu}) v_{H_{f}}
(\vec{P_{f}}) \right] \left( \frac{-ig^{\mu \nu}}{k^{2}} \right)
\left[ \bar{u}_{{h}_{f}} (\vec{p}_{f}) (-iqe \gamma_{\mu}) u_{h_{i}}
(\vec{p}_{i}) \right] , \label{feyn}
\end{eqnarray}
with $Qqe^{2}\/$ replaced by $g_{s}^{2}$ for the gluon-mediated diagram.
Note the use of helicity rather than spin eigenstate spinors, which is done in
order to implement a relativistic description of the mesons.  In a
nonrelativistic picture in which meson spin originates solely from the spin of
the quarks ({\em s}-waves), spin-0(1) mesons have spin-space wavefunctions
described by the usual singlet and triplet quark wavefunction $\bar{Q} q$
combinations:
\begin{eqnarray}
\frac{
\bar{Q}_{\uparrow} q_{\downarrow} \pm \bar{Q}_{\downarrow} q_{\uparrow}}
{\sqrt{2}}, & & \uparrow , \downarrow {\rm spins.}
\end{eqnarray}
The above expression remains true in a relativistic picture if we take the
initial and final spin-quantization axes to coincide with the axes of relative
momenta $\vec{p}$ and $\vec{p} \, ',$ respectively, and then take $\uparrow$,
$\downarrow$ as {\em helicity\/} eigenstates.  This is nothing more than the
simplest nontrivial case of the Jacob-Wick formalism~\cite{Jacob}.  It is
then a simple matter to write the constraint function for singlet (triplet)
mesons:
\begin{equation}
f(helicities)= \frac{1}{\sqrt{2}} (\delta_{h_{i} \uparrow} \delta_{H_{i}
\downarrow} \pm \delta_{h_{i} \downarrow} \delta_{H_{i} \uparrow}) \,
\frac{1}{\sqrt{2}} (\delta_{h_{f} \uparrow} \delta_{H_{f} \downarrow}  \pm
\delta_{h_{f} \downarrow} \delta_{H_{f} \uparrow}) , \label{const}
\end{equation}
and so the object of interest is the constrained matrix element ${\cal
M}_{sing}$ or ${\cal M}_{trip}$, which is the Feynman amplitude
multiplied by the constraint function and summed over spins (or helicities).
This is the object that is Fourier transformed in Eq.~\ref{deltae}.

In summary, mass contributions due to a binding interaction in a system of
particles may be computed by writing down the Feynman amplitude induced by the
interaction Hamiltonian, constraining the component particles to satisfy the
symmetry properties of the system, and convolving with the appropriate system
wavefunction.  The specific implementation of this technique to spin-0 and
spin-1 mesons with constituent quarks in a relative $\ell =0$ state is
described by Eqs.~\ref{deltae}, \ref{feyn}, and \ref{const}.

\section{The Nonrelativistic Limit}
With the method for computing mass contributions in hand, we find ourselves
with
two possible courses of action.  The first is to compute
${\cal M}_{sing}$ or ${\cal M}_{trip}$
in a fully relativistic manner, and then Fourier transform the result to obtain
$\Delta E_{CM}$.  The second is to immediately reduce the spinor bilinears via
Pauli approximants, thus producing a nonrelativistic expansion.  Let us explore
both directions for the pseudoscalar case; the vector case is not much
different.

The relativistic result is noncovariant, because the energy contribution is
evaluated specifically in the CM frame of the quarks.  We see this reflected
in the computation of the matrix element.  For example, it is convenient to
eliminate spinors from the calculation by means of relations like
\begin{equation}
\sum_{h} u_{h} (\vec{p}_{A}) \bar{u}_{h} (\vec{p}_{B}) = \frac{(m_{A}+p
\hspace{-1ex} /_{A})}
{\sqrt{2 m_{A} (E_{A}+m_{A})}} \, \frac{(1 + \gamma_{0})}{2} \,
\frac{(m_{B}+p \hspace{-1ex} /_{B})}{\sqrt{2 m_{B} (E_{B}+m_{B})}},
\end{equation}
and the explicit $\gamma_{0}$ is a signal of the noncovariance.  Once the
spinor
reductions and the resultant trace are performed, we find the expression
\begin{equation}
{\cal M}_{sing} = -(Qqe^{2} + g_{s}^{2}) {\cal NT} \frac{1}{k^{2}},
\end{equation}
where $\cal N$ results from the normalization factors, and $\cal T$ is the
gamma-matrix trace.  They are given by
\begin{eqnarray}
{\cal N} & = & \frac{1}{(2 \pi)^{6}} \frac{1}{2^{5}} \left[ E_{i} (E_{i} + M)
E_{f} (E_{f} + M) \e_{i} (\e_{i} + m) \e_{f} (\e_{f} + m) \right]^{-1/2}
\nonumber \\
{\rm and} & & \nonumber \\
{\cal T} & = & 8 \left\{ (p_{i} \cdot P_{i}) \left[ 2 \e_{f} E_{f} + 3(m E_{f}
+
M \e_{f} + mM) \right] \right. \nonumber \\
& & + (p_{f} \cdot P_{f}) \left[ 2 \e_{i} E_{i} + 3(m E_{i} + M \e_{i} +mM)
\right] + (p_{i} \cdot P_{i}) (p_{f} \cdot P_{f}) \nonumber \\
& & -(p_{i} \cdot p_{f}) \left[ 2 E_{i} E_{f} + M (E_{i} + E_{f} + M) \right]
\nonumber \\
& & -(P_{i} \cdot P_{f}) \left[ 2 \e_{i} \e_{f} + m (\e_{i} + \e_{f} + m)
\right] + (p_{i} \cdot p_{f}) (P_{i} \cdot P_{f}) \nonumber \\
& & -(p_{i} \cdot P_{f}) \left[ m E_{i} + M \e_{f} + mM \right] \nonumber \\
& & -(P_{i} \cdot p_{f}) \left[ m E_{f} + M \e_{i} + mM \right]
-(p_{i} \cdot P_{f}) (P_{i} \cdot p_{f}) \nonumber \\
& & + \left[ -2mM(E_{i} - E_{f})(\e_{i} - \e_{f}) \right. \nonumber \\
& & +mM \left( m(E_{i} + E_{f}) + M(\e_{i} + \e_{f}) + mM \right) \nonumber \\
& & \left. \left.+ 2 m^{2} E_{i} E_{f} + 2 M^{2} \e_{i} \e_{f} \right]
\right\}.
\end{eqnarray}
Also,
\begin{equation}
k^{2} = (p_{i} - p_{f})^{2} = (\e_{i} - \e_{f})^{2} - (\vec{p} - \vec{p} \, ')
^{2}.
\end{equation}
It is, in principle, possible to Fourier transform the product ${\cal
M}_{sing}$
of these unwieldy functions to obtain the full relativistic result for $\Delta
E_{CM}$; this has not yet been performed.  We can also perform the expansion of
the energy factors in powers of $\frac{p}{m}$, where all such
momentum-over-mass quotients that occur are taken to be of the same order.

However, this is unnecessary work, for if we require only a nonrelativistic
expansion, there is a much faster way, namely expansion of the spinor bilinears
via the Pauli approximants
\begin{eqnarray}
\bar{u} (\vec{p} \, ') \vec{\gamma} u(\vec{p}) & = & \bra{\chi '}
\frac{(\vec{p} + \vec{p} \, ')}{2m} + i \frac{\vec{\sigma}
\times (\vec{p} \, ' - \vec{p})}{2m} \ket{\chi}
+ o \left[ \left( \frac{p}{m} \right) ^{3} \right] \nonumber \\
\bar{u} (\vec{p} \, ') \gamma^{0} u(\vec{p}) & = & \bra{\chi '} 1 +
\frac{(\vec{p} + \vec{p} \, ')^{2}}{8m^{2}} + i \frac{\vec{\sigma} \cdot
(\vec{p} \, ' \times \vec{p})}{4m^{2}} \ket{\chi} + o \left[ \left( \frac{p}{m}
\right)^{4} \right] .
\end{eqnarray}
Using these expansions in Eq.~\ref{feyn} and taking $\ket{\chi} , \ket{\chi '}$
in helicity basis, we quickly find
\begin{eqnarray}
{\cal M}_{sing} & \stackrel{NR}{\Rightarrow} & \frac{Qqe^{2}+g_{s}^{2}}
{(2 \pi)^{6}(\vec{p} - \vec{p} \, ')^{2}}
\left\{ 1 + \frac{(\vec{p} + \vec{p} \, ')^{2}}{4mM}
- \frac{(\vec{p} - \vec{p} \, ')^{2}}{8} \left[ \frac{1}{m^{2}} -
\frac{4}{mM} + \frac{1}{M^{2}} \right] \right. \nonumber \\
& & \left. + o \left[ \left( \frac{p}{m} \right)^{4} \right] \right\} .
\end{eqnarray}

The gluon diagram has the additional physical constraint that the initial and
final $q \bar{q}\/$ pairs are combined into a color singlet; this introduces an
additional factor of $-\frac{4}{3}\/$.  Then Fourier transformation of this
result produces
\begin{eqnarray}
\Delta E_{CM, sing} & = & \label{sing} \left( \alpha Qq - \frac{4}{3}
\alpha_{s} \right) \nonumber \\
& & \left\{ \vev{\frac{1}{r}} + \frac{1}{2mM} \vev{\frac{1}{r}
(\vec{p} \, ^{2} + \hat{r} \cdot (\hat{r} \cdot \vec{p} \, ) \vec{p} \, )}
\right. \nonumber \\
& & \left. -\frac{\pi}{2} \left( \frac{1}{m^{2}} -\frac{4}{mM} +
\frac{1}{M^{2}}
\right) \vev{\delta^{3} (\vec{r} \, )} \right\} + \cdots
\end{eqnarray}
In comparison, the energy contribution from the Breit-Fermi interaction
(Eq.~\ref{bf}) for a quark-antiquark pair of masses {\em m, M\/} in the CM
reduces to
\begin{eqnarray}
\lefteqn{\vev{H_{BF}} = \left( \alpha Qq - \frac{4}{3} \alpha_{s} \right)} & &
\nonumber \\ \label{vev}
& & \left\{ \vev{\frac{1}{r}} + \frac{1}{2mM} \vev{\frac{1}{r} (\vec{p} \, ^{2}
+ \hat{r} \cdot (\hat{r} \cdot \vec{p} \, ) \vec{p} \, )} \right. \nonumber \\
& & -\frac{\pi}{2} \vev{\delta^{3} (\vec{r} \, )} \left[ \frac{1}{m^{2}} +
\frac{1}{M^{2}} + \frac{4}{mM} \left( {\cal G} + \delta_{S,1} \delta_{q
flavors}
\right) \right] \nonumber \\
& & \left. -\frac{1}{2} \vev{\frac{1}{r^{3}}} \vev{\vec{L} \cdot \left(
\frac{\vec{s}_{q}}{m^{2}} + \frac{\vec{s}_{\bar{Q}}}{M^{2}} + \frac{2
\vec{S}}{mM} \right) + \frac{S_{12}}{2mM} } \right\}
\end{eqnarray}
where ${\cal G} \equiv \frac{4}{3} \vev{\vec{s}_{q} \cdot \vec{s}_{\bar{Q}}}$,
which is $-1 \left( \frac{1}{3} \right)$ for $S = 0 (1)$.  Also, $\vec{S}
\equiv \vec{s}_{q} + \vec{s}_{\bar{Q}}$, and $S_{12}\/$ is the $\Delta
L = 2\/$ tensor operator
\begin{equation}
S_{12} \equiv 3 (\vec{\sigma}_{1} \cdot \hat{r})(\vec{\sigma}_{2} \cdot
\hat{r})
- \vec{\sigma}_{1} \cdot \vec{\sigma}_{2}.
\end{equation}
For mesons with differently-flavored quarks in a relative $\ell
=0\/$ state, many of the terms drop out.  Let us define
\begin{eqnarray}
B & \equiv & \vev{\frac{1}{r}} , \nonumber \\
C & \equiv & \vev{\frac{1}{r} (\vec{p} \, ^{2} + \hat{r} \cdot (\hat{r} \cdot
\vec{p} \, ) \vec{p} \, )} , \nonumber \\
D & \equiv & \vev{\delta^{3} (\vec{r} \, )} .
\end{eqnarray}
Then Eq.~\ref{vev} becomes
\begin{equation}
\vev{H_{BF}} = \left( \alpha Qq -\frac{4}{3} \alpha_{s} \right) \left[ B +
\frac{1}{2mM} C -\frac{\pi}{2} \left( \frac{1}{m^{2}} + \frac{1}{M^{2}} +
\frac{4{\cal G}}{mM}
\right) D \right] , \label{hbf}
\end{equation}
and this is exactly Eq.~\ref{sing} where ${\cal G}= -1$.

We have been up to now considering only the contributions to the mass
originating from the binding interaction due to one-gluon and one-photon
exchanges; there are, of course, also contributions from the kinetic energy of
the quarks.  Were we calculating these quantities in a relativistic theory, we
would simply compute
$K \! E =\vev{\sqrt{m^{2}+\vec{p} \, ^{2}}}$.  The square root may be formally
expanded in norelativistic quantum mechanics as well, resulting in an
alternating series in $\vev{\vec{p} \, ^{2n}}$.  However, for large enough
$n\/$ in NRQM, these
expectation values tend to diverge.  For example, in the hydrogen atom,
divergence occurs for $s$-waves at $n=3$.  Furthermore, if the system is not
highly nonrelativistic, the inclusion of the $\vev{\vec{p} \, ^{4}}$ may cause
us to grossly underestimate the true value of the kinetic energy.  The problem
is that there is no positive $\vev{\vec{p} \, ^{6}}$ term to balance the large
negative $\vev{\vec{p} \, ^{4}}$ term.  For these reasons, we incorporate the
alternating nature of the series in a computationally simple way by making the
ansatz
\begin{equation}
K \! E = \sqrt{m^{2} + \vev{\vec{p} \, ^{2}}} . \label{ans}
\end{equation}

In order to evaluate the expectation values in the above equations, we will
need
to choose a potential.  In the meantime, let us simply denote it with $U(r)$.
Then at last we have the mass formula:
\begin{equation}
M_{meson} = \sqrt{M^{2} + \vev{\vec{p} \, ^{2}}} + \sqrt{m^{2} +
\vev{\vec{p} \, ^{2}}} + \vev{U(r)} + \vev{H_{BF}} . \label{mass}
\end{equation}
The static potential $U(r)$ takes the place of $L$, the universal quark binding
function, in Eq. 1 of Ref~\cite{DeRujula}.

\section{Mass Splitting Formulae}
The static potential in which the quarks interact determines the form of the
NRQM wavefunction.  The strong Coulombic term gives the largest energy
contribution of terms within the Breit-Fermi interaction, and therefore would
also be expected to substantially alter the wavefunction in perturbation
theory.
Therefore, we include the strong Coulombic term in the static potential:
\begin{equation}
V(r) \equiv U(r) -\frac{4}{3} \frac{\alpha_{s}}{r}.
\end{equation}
Then the mass formula Eq.~\ref{mass} becomes, using Eq.~\ref{hbf},
\begin{eqnarray}
M_{meson} & = & \sqrt{M^{2} + \vev{\vec{p} \, ^{2}}} + \sqrt{m^{2} +
\vev{\vec{p} \, ^{2}}} + \vev{V(r)} +\alpha QqB \nonumber \\
& & + \left( \alpha Qq -\frac{4}{3} \alpha_{s} \right) \left[
\frac{1}{2mM} C -\frac{\pi}{2} \left( \frac{1}{m^{2}} + \frac{1}{M^{2}} +
\frac{4 {\cal G}}{mM} \right) D \right].
\end{eqnarray}

Now at last we are in a position to write explicit formulae for the mass
splittings of interest.  Denoting the mass of a meson of spin $S\/$ and
valence quarks $\bar{Q}$,$q\/$ as $M^{S}(\bar{Q}q)$, we define:
\begin{eqnarray}
\label{diff}
\Delta^{0}_{Q}  & \equiv & M^{0} (\bar{Q}u) - M^{0} (\bar{Q}d) \nonumber \\
\Delta^{1}_{Q}  & \equiv & M^{1} (\bar{Q}u) - M^{1} (\bar{Q}d) \nonumber \\
\Delta^{*}_{Qu} & \equiv & M^{1} (\bar{Q}u) - M^{0} (\bar{Q}u) \nonumber \\
\Delta^{*}_{Qd} & \equiv & M^{1} (\bar{Q}d) - M^{0} (\bar{Q}d),
\end{eqnarray}
where $u\/$ and $d,$ the up and down constituent quarks, are nearly
degenerate in mass: defining $\Delta m \equiv m_{u} - m_{d}\/$ and $m \equiv
\frac{m_{u} + m_{d}}{2}$, we have $\left| \frac{\Delta m}{m} \right| \ll 1$.
Therefore, the differences in Eq.~\ref{diff} are expanded in Taylor series in
$\frac{\Delta m}{m}$ about $m$.  It is also convenient to define
\begin{eqnarray}
        A      & \equiv & \vev{\vec{p} \, ^{2}} ,     \nonumber \\
    \beta      & \equiv & \frac{1}{1 + m/M} ,         \nonumber \\
      \mu      & \equiv & \mbox{usual reduced mass} , \nonumber \\
\bar{\mu}      & \equiv & m \beta ,                   \nonumber \\
D_{\alpha_{s}} & \equiv & \left. \beta \left( \frac{\mu}{\alpha_{s}}
\frac{\partial \alpha_{s}}{\partial \mu} \right) \right| _{\mu = \bar{\mu}},
\nonumber \\
    D_{X}      & \equiv & \left. \beta \left( \mu \frac{\partial X}{\partial
\mu} \right) \right| _{\mu = \bar{\mu}} , \; \; \; X=A,B,C,D,\vev{V(r)}.
\end{eqnarray}
Then the expressions for mass splitting are
\begin{eqnarray}
\label{d0}
\Delta_{Q}^{0,1} & = & \left[ \frac{2m^{2} +D_{A}}{\sqrt{m^{2} + A}} +
\frac{D_{A}}{\sqrt{M^{2} + A}} \right] \frac{\Delta m}{2m} + D_{\vev{V}}
\frac{\Delta m}{m} \nonumber \\
& & -\frac{4}{3} \alpha_{s} \Delta m \left\{ \frac{1}{2m^{2}M} \left(
D_{C} - C + CD_{\alpha_{s}} \right) \right. \nonumber \\
& & \left. -\frac{\pi}{2m^{3}} \left[ \left( 1 + 4 {\cal G} \frac{m}{M}
+ \frac{m^{2}}{M^{2}} \right) \left( D_{D} + DD_{\alpha_{s}}
\right) - 2 \left( 1 + 2 {\cal G} \frac{m}{M} \right) D \right] \right\}
\nonumber \\
& & + \alpha Q \left[ B + \frac{1}{2mM} C -\frac{\pi}{2m^{2}} \left(1 + 4
{\cal G} \frac{m}{M} + \frac{m^{2}}{M^{2}} \right) D \right] \nonumber \\
& & + o \left[ \left( \frac{\Delta m}{m} \right) ^{3} \right] + o \left( \alpha
\frac{\Delta m}{m} \right) .
\end{eqnarray}
Note that no derivatives appear in the $\alpha_{EM}$ terms because we take both
$\alpha_{EM}$ and $\frac{\Delta m}{m}$ (but not $\alpha_{s}$) as expansion
parameters.  Furthermore, the running of $\alpha_{s} (\mu)$ is explicitly taken
into account.

For vector-pseudoscalar splittings, we have
\begin{eqnarray}
\label{d1}
\Delta_{Qq}^{*} & = & \frac{8 \pi}{3mM} \left[ \left( \frac{4}{3} \alpha_{s} -
\alpha Qq \right) D \pm \frac{4}{3} \alpha_{s} \frac{\Delta m}{2m} \left( D_{D}
+ DD_{\alpha_{s}} - D \right) \right] \nonumber \\
& & + o \left[ \left( \frac{\Delta m}{m} \right) ^{2} \right] + o \left( \alpha
\frac{\Delta m}{m} \right) , \; \; \; \; \mbox{with } \pm \; \mbox{for }
q=u(d).
\end{eqnarray}
Let us remind ourselves of the physical significance of the terms in the
previous two equations.  Terms containing $A\/$ signify kinetic energy
contributions, including intrinsic quark masses.  The potential term is
identified, of course, by $V$; $B$, $C$, and $D$ denote static Coulomb, Darwin,
and hyperfine terms, respectively.

\section{Quantum-mechanical Theorems}
In order to apply the foregoing results, we will need to evaluate the
expectation values {\em A,B,C,D,} and $\vev{V(r)}$ for our potential $V(r)$.
There are two quantum-mechanical theorems which make the evaluation of these
expectation values and their mass derivatives simpler \cite{Quigg}.  The first
is the
\begin{fh}[Feynman-Hellmann Theorem]
For normalized eigenstates of a Hamiltonian depending on a parameter $\lambda$,
\begin{equation}
\frac{\partial E}{\partial \lambda} = \vev{ \frac{\partial H(\lambda)}
{\partial
\lambda}} .
\end{equation}
In the particular case that $\lambda = \mu$,
\begin{equation}
\frac{\partial E}{\partial \mu} = -\frac{1}{\mu} \left( E - \vev{V(r)} \right)
+
\vev{\frac{\partial V}{\partial \mu}}  . \label{fhapp}
\end{equation}
\end{fh}

The other result may be less familiar.  For reasons that will become clear, let
us call it the
\begin{gvth}[Generalized Virial Theorem]
Consider bound eigenstates \\
$u_{\ell} (r)$ in a spherically symmetric potential V(r) such that
\begin{displaymath}
\lim_{r \rightarrow 0} \, r^{2} V(r) = 0 .
\end{displaymath}
Then, writing the Schr\"{o}dinger equation as
\begin{displaymath}
u^{''}_{\ell} (r) + \frac{2\mu}{\hbar ^{2}} \left[ E - V(r) -\frac{\hbar^{2}
\ell (\ell +1)}{2 \mu r^{2}} \right] u_{\ell} (r) = 0 ,
\end{displaymath}
and defining $a_{\ell}$ by
\begin{displaymath}
\lim_{r \rightarrow 0} \frac{u_{\ell}(r)}{r^{l+1}} \equiv a_{\ell} ,
\end{displaymath}
then \\
\mbox{ i)} $a_{\ell}$ is a nonzero constant;
\vspace{2ex} \\
\mbox{ii)} for $q \geq -2\ell$,
\begin{eqnarray}
(2 \ell + 1)^{2} a_{\ell}^{2} \delta_{q,-2l} & = & -\frac{2 \mu}{\hbar ^{2}}
\vev{r^{q-1} \left( 2q(E-V(r)) -r \frac{dV}{dr} \right) } \nonumber \\
& & + (q-1) \left[ 2\ell (\ell +1) -\frac{1}{2} q(q-2) \right] \vev{r^{q-3}} .
\end{eqnarray}
\end{gvth}

Clearly this theorem will prove most useful for potentials easily expressed as
a
sum of terms which are powers in $r$.  But in fact there are some interesting
general results included.  For example, the $q=\ell =0$ case generates the
well-known result for $s$-waves,
\begin{equation}
\left| \Psi (0) \right| ^{2} = \frac{\mu}{2 \pi \hbar ^{2}}
\vev{\frac{dV}{dr}},
\end{equation}
whereas the $q=1$ case produces
\begin{equation}
E - \vev{V(r)} = \frac{1}{2} \vev{ r \frac{dV}{dr}},
\end{equation}
the quantum-mechanical virial theorem.

Using partial integration, the Schr\"{o}dinger equation, and the GVT, it is
possible to show the following ($\hbar = 1$):
\begin{eqnarray}
A & = & 2 \mu \left( E - \vev{V(r)} \right) , \nonumber \\
C & = & 4 \mu \left[ E \vev{\frac{1}{r}} - \vev{\frac{V(r)}{r}} -\frac{1}{4}
\vev{\frac{dV}{dr}} \left( 1 + \delta_{\ell , 0} \right) \right] , \nonumber \\
D & = & \frac{\mu}{2 \pi} \vev{\frac{dV}{dr}} \delta_{\ell , 0} , \nonumber \\
\int_{0}^{\infty} \left( \frac{du_{\ell}(r)}{dr} \right) ^{2}  dr & = & A -
\ell (\ell + 1) \vev{\frac{1}{r^{2}}} .
\end{eqnarray}

In addition, we must also uncover what we can about the $\mu$-dependence of
expectation values.  For a general potential this is actually an unsolved
problem.  However, unless the potential has very special $\mu$-dependence, it
can be shown that only in the case $V(r) = V_{0} r^{\nu}$ is it possible to
scale away the dimensionful parameters $V_{0}$ and $\mu$ in the Schr\"{o}dinger
equation.  In that case, the $\mu$-dependence will be entirely contained in the
scaling factors, and computing $D_{X}$ will be trivial.  Unfortunately, in the
potential we consider in the next section, we will see that this is not the
case, and we must resort to subterfuge to obtain the required information.

\section{Example: $V(r)=\frac{r}{a^{2}}-\frac{\kappa}{r}$}
The potential $V(r) = \frac{r}{a^{2}} -\frac{\kappa}{r}$, where $\kappa =
\frac{4}{3} \alpha_{s}$, is interesting because it phenomenologically includes
quark confinement via the linear term.  This potential was considered in
greatest detail by Eichten {\em et al.}\ \cite{Eichten} to describe the mass
splitting structure of the charmonium system (and was later applied to
bottomonium).  The Schr\"{o}dinger equation was solved numerically; currently,
no analytic solution is known.  However, it is possible to extract a great deal
of information from their tabulated results, as we shall see below.

This is possible because of the GVT.  If we rescale the Schr\"{o}dinger
equation with the linear-plus-Coulomb potential to
\begin{equation}
\left( \frac{d^{2}}{d \rho^{2}} - \frac{\ell (\ell + 1)}{\rho^{2}} +
\frac{\lambda}{\rho} +\zeta -\rho \right) w_{\ell} (\rho) = 0 ,
\end{equation}
where
\begin{eqnarray}
\rho \equiv \left( \frac{2 \mu}{a^{2}} \right) ^{1/3} r , & & \lambda \equiv
\kappa (2 \mu a)^{2/3} , \nonumber \\
\zeta \equiv (2 \mu a^{4})^{1/3} E , & & w_{\ell}(\rho) \equiv u_{\ell}(r)
\left( \frac{a^{2}}{2 \mu} \right) ^{1/6} , \label{scale}
\end{eqnarray}
then the GVT gives
\begin{eqnarray}
(q=0) \; \; \; a_{0}^{2} \, \delta_{0,\ell} & = & \left( \frac{2 \mu}
{a^{2}} \right) \left[ 1+ \lambda \vev{\frac{1}{\rho^{2}}} -2 \ell (\ell + 1)
\vev{\frac{1}{\rho^{3}}} \right], \nonumber \\
(q=1) \; \; \; \; \; \; \; \; \; \;  0 & = & 3 \vev{\rho} -2 \zeta -\lambda
\vev{\frac{1}{\rho}} .
\end{eqnarray}
Also, defining
\begin{equation}
\vev{v^{2}} \equiv \int_{0}^{\infty} \left( \frac{dw_{\ell}(\rho)}{d \rho}
\right) ^{2} d \rho ,
\end{equation}
we find
\begin{equation}
\vev{v^{2}} = -\vev{\rho} + \zeta + \lambda \vev{\frac{1}{\rho}} -\ell (\ell +
1) \vev{\frac{1}{\rho^{2}}} .
\end{equation}

It is a happy accident of this potential that all of the quantities in the
expectation values we need, for any $\ell$, may be expressed in terms of the
three quantities $\zeta , \: \vev{\frac{1}{\rho^{2}}}$, and $\vev{v^{2}}$,
which are exactly those values tabulated for the $1s$-state, as functions of
$\lambda$, in Eichten {\em et al.\/} Table I.  Defining $\sigma \equiv \left(
\frac{2 \mu}{a^{2}} \right)^{1/3}$ and taking $\ell = 0$ (as per our mesonic
model), we find
\begin{eqnarray}
A & = & \sigma^{2} \vev{v^{2}} , \nonumber \\
B & = & \frac{\sigma}{2 \lambda} \left[ 3 \vev{v^{2}} -\zeta \right] ,
\nonumber
\\
C & = & \sigma^{2} \left[ 2B \zeta + \sigma \left( -3 + \lambda
\vev{\frac{1}{\rho^{2}}} \right) \right] , \nonumber \\
D & = & \frac{\sigma^{3}}{4 \pi} \left[ \lambda \vev{\frac{1}{\rho^{2}}} + 1
\right] . \label{dee}
\end{eqnarray}
So now we can compute all of the necessary expectation values numerically.  The
superficial singularity in $B(\lambda = 0)$ is false; $B(0)$ is computed by
extrapolation of the computed values of $B$ for nonzero $\lambda$ and is found
to be finite.

The mass derivatives must be handled in a different fashion.  We begin by
defining
\begin{eqnarray*}
\tilde{D}_{\zeta} \equiv \mu \frac{\partial \zeta}{\partial \mu} , &
\tilde{D}_{v} \equiv \mu {\displaystyle \frac{\partial \vev{v^{2}}}{\partial
\mu}} , &
\tilde{D}_{\rho} \equiv \mu \frac{\partial \vev{\frac{1}{\rho^{2}}}}{\partial
\mu} ,
\end{eqnarray*}
and
\begin{equation}
\tilde{D}_{\alpha_{s}} \equiv \frac{\mu}{\alpha_{s}} \frac{\partial
\alpha_{s}}{\partial \mu} .
\end{equation}
{}From the Feynman-Hellmann theorem (Eq.~\ref{fhapp}) we may show
\begin{equation}
\tilde{D}_{\zeta} = \left( \frac{\zeta}{3} - \vev{v^{2}} \right) \left( 1 +
\frac{3}{2} \tilde{D}_{\alpha_{s}} \right) .
\end{equation}
As mentioned in the previous section, scaling of the Schr\"{o}dinger equation
can be accomplished for $\mu$-independent potentials that are monomials.  In
the case $\lambda = 0$ (a purely linear potential), the scaling would be
perfect, and $\zeta, \, \vev{\frac{1}{\rho^{2}}},$ and
$\vev{v^{2}}$ would be $\mu$-independent.  In the $\lambda \neq 0$ case, the
derivatives must be found numerically.  Again, fortunately, we have a table of
numerical values of the desired expectation values, as a function of $\lambda
(\mu)$.  We fit the expectation values $Y$ ($ = \vev{\frac{1}
{\rho^{2}}}, \, \vev{v^{2}}$) to the functional form
\begin{equation}
Y(\lambda) = Y_{0} + K \lambda^{n_{Y}} .
\end{equation}
Then, using Eq.~\ref{scale}, we find
\begin{equation}
\tilde{D}_{Y} = \left( \frac{2}{3} + \tilde{D}_{\alpha_{s}} \right) n_{Y}
\left( Y - Y_{0} \right) .
\end{equation}
Finally, define
\begin{equation}
\tilde{D}_{X} \equiv \mu \frac{\partial X}{\partial \mu} \mbox{ for }
X=A,B,C,D,
\end{equation}
so that
\begin{equation}
D_{X} = \beta \left. \tilde{D}_{X} \right| _{\mu = \bar{\mu}} .
\end{equation}
Then we find
\begin{eqnarray}
\tilde{D}_{A} & = & \frac{2}{3} A + \sigma^{2} \tilde{D}_{v} , \nonumber \\
\tilde{D}_{B} & = & \frac{3 \sigma}{2 \lambda} \tilde{D}_{v} - \frac{1}{2} B
\tilde{D}_{\alpha_{s}} , \; \; \; ( \lambda \neq 0), \nonumber \\
\tilde{D}_{C} & = & \frac{5}{3} C + 2 \sigma^{2} \left\{ \left( -\zeta +
\tilde{D}_{\zeta} \right) B + \zeta \tilde{D}_{B} + \sigma \left[
\frac{\lambda}{2} \left( \tilde{D}_{\rho} + \tilde{D}_{\alpha_{s}}
\vev{\frac{1}{\rho^{2}}} \right) + 1 \right] \right\} , \nonumber \\
\tilde{D}_{D} & = & \frac{\sigma^{3}}{4 \pi} \left\{ \lambda \left[ \left(
\frac{5}{3} + \tilde{D}_{\alpha_{s}} \right)
\vev{\frac{1}{\rho^{2}}} + \tilde{D}_{\rho} \right] + 1 \right\} .
\end{eqnarray}
In the exceptional case of $\tilde{D}_{B}$, we simply note that, for $\lambda =
0$, we have perfect scaling of the wave equation, and we can quickly show that
$\left. \tilde{D}_{B} \right| _{\lambda = 0} = \left. \frac{1}{3} B \right|
_{\lambda = 0}$.  This provides us with everything we need to produce numerical
results.

Before leaving the topic, let us mention that many complications of
$\mu$-derivatives of expectation values vanish if the potential itself has the
appropriate $\mu$-dependence, for then scaling of the wave equation is
possible.
For example, one can scale the Schr\"{o}dinger equation for the potential
\begin{equation}
V(r) = c \mu^{2} r - \frac{\kappa}{r} ,
\end{equation}
where $c\/$ is a pure number.

\section{Numerical Results}
The method of obtaining results from the theory requires us to choose several
numerical inputs, most of which are believed known to within a few percent.
Let
us choose the following inputs to the model:
\begin{eqnarray}
m = 340 \mbox{ MeV,} & M_{s} = 540 \mbox{ MeV,} & \nonumber \\
M_{c} = 1850 \mbox{ MeV,} & M_{b} = 5200 \mbox{ MeV,} & a = 1.95 \mbox{ GeV}
^{-1}. \label{nums}
\end{eqnarray}
The light quark constituent mass is arrived at by assuming that nucleons
consist of quarks with negligible anomalous magnetic moments, which can be
added
nonrelativistically to provide the full nucleonic magnetic moment.  Likewise,
the strange quark mass issues from the same considerations applied to strange
baryons \cite{DeRujula}.  The $c$ and $b$ quark masses are simply found by
dividing the threshold energy value for charm and bottom mesons by two
(however, smaller masses have been predicted using semileptonic decay
results in addition to meson masses \cite{Olsson}).  The confinement constant
is
inferred from charmonium levels \cite{Eichten}.

One important input not yet mentioned is $\Delta m$, the up-down quark mass
difference.  Traditionally, this assumes a value of $\approx -3$ to $-8\/$ MeV,
in order to account for the electromagnetic mass splittings of the lighter
hadrons.  In this model, with the inputs listed in Eq.~\ref{nums}, we find that
the experimental splittings for the $D\/$- and $B\/$-mesons (both vector and
pseudoscalar) can be satisfied within one standard deviation of experimental
error for values of $\Delta m\/$ in the narrow range of $-4.05$ to $-4.10$
MeV\@.  In contrast, it is found that for no choice of $\Delta m\/$ can one
simultaneously fit $D$- and $K$-meson data simultaneously, as was done in the
earlier models.

Before exhibiting the quantitative results, let us describe the method by which
they are obtained.  Once particular inputs for the above variables are chosen,
one can compute the various mass splittings for the values of $\lambda \propto
\alpha_{s}\/$ that occur in Table I of Ref.~\cite{Eichten}, and in-between
values may be interpolated.  We then fit vector-pseudoscalar splittings
(computed via Eq.~\ref{d1}) to the corresponding experimental data (since these
numbers have the smallest relative errors of the splittings we consider)
and thus obtain a value of $\alpha_{s}$.  For the three systems $K$, $D$, and
$B$, we use the three values of $\alpha_{s}$ to estimate graphically (and
admittedly rather crudely) its mass derivative.
Applying the values of the strong coupling constant and its derivative
to the splittings in Eq.~\ref{d0}, we generate all of the other values.  If the
resultant numbers do not fall within the experimental error bars for such
splittings, we vary the input parameters (most importantly, $\Delta m$) until
a simultaneous fit is achieved.

Table I displays the various contributions to mass splittings derived in this
fashion for $B\/$- and $D\/$-mesons.  Although the kinetic term (which includes
the explicit difference $\Delta m$) and the static Coulomb term are
unsurprisingly large, a significant contribution to the mass splitting arises
in the strong hyperfine term.  That strong contributions to the so-called
electromagnetic mass splittings could be important was observed by
Chan~\cite{Chan}, and was exploited in the subsequent literature.  It is
exactly this term which is most significant in driving the $B$ splittings
toward zero.  Note also the decrease in the derived value of $\alpha_{s}\/$ as
the reduced mass of the system increases when we move from the $D\/$ system to
the $B\/$ system, consistent with asymptotic freedom in QCD.  It was this
running which motivated the inclusion of mass derivatives of the strong
coupling
constant in this model.  If they are not included, one actually obtains a value
of $\Delta m > 0$, in contrast with all estimates from both nonrelativistic and
chiral models.

The net result is that one can satisfactorily fit the data for the $D\/$ and
$B\/$ systems simultaneously in the most natural nonrelativistic model with a
physically reasonable potential.  The comparison of the results of this
calculation for $\Delta m = -4.10$ MeV to experimental data is presented in
Table II\@.

However, the table also exhibits very bad agreement for the $K\/$ system
(despite the fact that the fit to vector-pseudoscalar splittings yields the
value $\alpha_{s} = 0.424$, which runs in the correct direction).  One
may view this as a failure of the nonrelativistic assumptions of the model in
a variety of ways.  Most obvious are the ansatz Eq.~\ref{ans}, which is
certainly not an airtight assumption in even the best of circumstances, and the
crudeness of the estimate of $\frac{\partial \alpha_{s}}{\partial \mu}$.  Other
possible problems include the assumption that the quarks occur only in a
relative $\ell=0$ state (relevant for $K^{*}$-mesons), and the assumption that
the strong effects are dominated by a confining potential and one-gluon
exchange, since at the lower energies associated with the $K\/$ system,
$o(\alpha_{s}^{2})$ terms and more complex models of confinement may be
required.  The failure of these assumptions can drastically alter the strong
hyperfine interaction, which determines the size of $\alpha_{s}$, and hence
the other mass splittings.

Some may find the small size of $\alpha_{s}\/$ somewhat puzzling.  This is
primarily the result of the confining term of the model potential: it causes
the
wavefunction to be large at the origin, and thus a small $\alpha_{s}$ is
required to give the same experimentally measured vector-pseudoscalar splitting
(see Eq.~\ref{d1}).  Such small values for the strong coupling constant might
lead to excessively small values of $\Lambda_{QCD}\/$ and large values for
mesonic decay constants $f_{\bar{Q}q}$.  Indeed, given the naive expressions
for these quantities:
\begin{equation}
\alpha_{s}(\mu) = \frac{12 \pi}{(33 - 2 n_{f}) \log \left(\frac{\mu^{2}}
{\Lambda^{2}} \right) } ,
\end{equation}
and, assuming the relative momenta of the quarks is small,
\begin{equation}
f_{\bar{Q}q}^{2} = \frac{12}{M_{\bar{Q}} + m_{q}} \left| \Psi (0) \right| ^{2},
\end{equation}
let us consider, for example, the $D\/$ system. Then $\alpha_{s} = 0.363$ and
$\mu = 287$ MeV, and with three flavors of quark, we calculate $\Lambda_{QCD} =
42$ MeV and $f_{B} = 342$ MeV\@.  However, one may state the following
objections: first, $\Lambda_{QCD}$ is computed from the full theory of QCD, but
the nonrelativistic potential approach includes the confinement in an
{\em ad hoc\/} fashion, by including a confinement constant $a$, which is
independent of $\alpha_{s}$.  Furthermore, choosing $\Lambda_{QCD}$ as the
renormalization point forces an artificial singularity at $\mu =
\Lambda_{QCD}$;
the problem is that little is known about the low-energy behavior of strong
interactions.  At low energies the computation and interpretation
of $\Lambda_{QCD}$ requires a more careful consideration of confinement.
With respect to the decay constant, the assumption that the quarks are
relatively at rest leads to the evaluation of the wavefunction at zero
separation.  Inclusion of nonzero relative momentum will presumably result in
the necessity of considering separations of up to a Compton wavelength
$r \approx \frac{1}{\mu}$, for which the wavefunction is smaller in the
1$s$-state.  Thus decay constants may be smaller than computed in the naive
model.

There is one further qualitative success of this model, a partial explanation
of the experimental facts that $D^{*}_{s} - D_{s} = 141.5 \pm 1.9$ MeV
$\approx D^{*} - D$, and $B^{*}_{s} - B_{s} = 47.0 \pm 2.6$ MeV
$\approx B^{*} - B$, namely, the approximate independence of
vector-pseudoscalar
splitting on the light quark mass.  In our model, the leading term of the
splitting is, using Eqs.~\ref{d1} and \ref{dee},
\begin{equation}
\Delta^{*}_{Qq} \approx \frac{16}{9Ma^{2}} \alpha_{s} \beta \left[ \lambda
\vev{\frac{1}{\rho^{2}}} +1 \right] .
\end{equation}
Inasmuch as $\beta$, $\lambda \vev{\frac{1}{\rho^{2}}}$, and $\alpha_{s}$ are
slowly varying in the light quark mass $m$, the full expression reflects this
insensitivity, in accord with experiment.  In fact, we may fit the experimental
values above to obtain more running values of $\alpha_{s}$:
\begin{eqnarray}
\Delta^{*}_{cs} = 141.5 \mbox{ MeV} & \mbox{for} & \alpha_{s} = 0.351 ,
\nonumber \\
\Delta^{*}_{bs} =  47.0 \mbox{ MeV} & \mbox{for} & \alpha_{s} = 0.295 ,
\end{eqnarray}
and again these decrease as the mass scale increases.  Note, however, one kink
in this interpretation: the heavy-strange mesons all have larger reduced masses
than their unflavored counterparts, yet the corresponding values of
$\alpha_{s}$
are nearly the same.

\section{Conclusions}
In this paper, we have seen how mass contributions to a bound system of
particles are derived from an interaction Hamiltonian in field theory, and how
this calculation is then reduced to a problem in nonrelativistic quantum
mechanics.  For the system of a quark and antiquark bound in a meson, the
exchange of one mediating vector boson reduces to the Breit-Fermi interaction
in the nonrelativistic limit.  It is also important to consider contributions
to
the total energy from the kinetic energy and the long-range potential of the
system; in fact, the higher-order momentum expectation values can be so large
that it is necessary to impose an ansatz (Eq.~\ref{ans}) in order to estimate
their combined effect.  Future work may suggest better estimates.

    It is found in the case of a linear-plus-Coulomb potential that the
largest contributions to electromagnetic mass splittings originate in the
kinetic energy, static Coulomb, and strong hyperfine terms.  However, it is
likely that similar results hold for other ans\"{a}tze and potentials.  As
in other models, vector-pseudoscalar mass differences are determined by strong
hyperfine terms.

    With typical values for quark masses, the confinement constant, and the
up-down quark mass difference, we can obtain agreement for the mass splittings
of the $D$- and $B$-mesons.  The failure of the model for $K\/$ mass splittings
is attributed to the collapse of the nonrelativistic assumptions in that case.
The model also qualitatively explains the similarity of heavy-strange to
heavy-unflavored vector-pseudoscalar splittings, although additional work is
needed to explain why these numbers are nearly equal, despite the expected
inequality of $\alpha_{s}$ at the two different energy scales.

    Another interesting problem is the running of $\alpha_{s}$ itself at low
energies.  As mentioned in the results section, this running cannot be
neglected
if we are to obtain sensible results, and yet our approximation of this running
is based on crude assumptions.  The size of $\alpha_{s}$ also enters into
another possible development, namely, whether terms of $o(\alpha_{s}^{2})$ are
important, particularly for the $K$ system.  More reliable estimates are
required.

    In addition to the explicit formulae derived in this paper, the techniques
employed here may be applied to later efforts: in particular, the explicit
consideration of the mass-dependence of expectation values and the use of
quantum-mechanical theorems to relate various expectation values for certain
potentials.  The methods and formulae in this work may prove to be a starting
point for subsequent research.

\section*{Acknowledgments}
I would like to thank J. D. Jackson, L. J. Hall, and M. Suzuki for
invaluable discussions during the preparation of this work.

\newpage
\renewcommand{\thetable}{\Roman{table}}
\begin{table}
\centering
\caption{Contributions to mass splittings of heavy mesons}
\begin{tabular}{||l|r|r||} \hline
                  & {\em D\/} mesons & {\em B\/} mesons \\ \hline
$\alpha_{s}$      &            0.363 &            0.312 \\ \hline
Source            &              MeV &              MeV \\ \hline\hline
Isospin pairs     &                  &                  \\ \hline
Kinetic energy    &           -4.109 &           -3.523 \\
Potential energy  &            1.057 &           -1.645 \\
Strong Darwin     &           -0.834 &           -0.635 \\
EM Darwin         &           -0.769 &            0.147 \\
Static Coulomb    &           -2.442 &            1.252 \\
                  &                  &                  \\
$\Delta^{0}_{Q}$  &                  &                  \\
Strong hyperfine  &            2.148 &            4.075 \\
EM hyperfine      &            0.424 &           -0.561 \\ \hline
{\em Total}       &           -4.525 &           -0.889 \\
                  &                  &                  \\
$\Delta^{1}_{Q}$  &                  &                  \\
Strong hyperfine  &            3.683 &            5.244 \\
EM hyperfine      &            1.817 &           -0.825 \\ \hline
{\em Total}       &           -1.596 &            0.017 \\ \hline\hline
$1^{-} - 0^{-}$   &                  &                  \\ \hline
Strong hyperfine  &                  &                  \\
       (leading)  &           141.30 &            46.04 \\
    (subleading)  &       $\pm$ 0.77 &       $\pm$ 0.58 \\
                  &                  &                  \\
$\Delta^{*}_{Qu}$ &                  &                  \\
EM hyperfine      &             0.93 &            -0.18 \\ \hline
{\em Total}       &           143.00 &            46.45 \\
                  &                  &                  \\
$\Delta^{*}_{Qd}$ &                  &                  \\
EM hyperfine      &            -0.46 &             0.09 \\ \hline
{\em Total}       &           140.07 &            45.54 \\ \hline
\end{tabular}
\end{table}

\newpage
\begin{table}
\centering
\caption{Meson mass splittings compared to experiment}
\begin{tabular}{||c|l|r|r||} \hline
Mass splitting    & Notation          & Predicted (MeV) & Expt. (MeV) \\ \hline
$K^{+} - K^{0}$   & $\Delta^{0}_{s}$  &           -0.98 & -4.024 $\pm$ 0.032 \\
$K^{*+} - K^{*0}$ & $\Delta^{1}_{s}$  &           -0.15 & -4.51 $\pm$ 0.37
$^{a}$ \\
$K^{*+} - K^{+}$  & $\Delta^{*}_{su}$ &           398.6 & 397.94 $\pm$ 0.24
$^{a}$ \\
$K^{*0} - K^{0}$  & $\Delta^{*}_{sd}$ &           397.8 & 398.43 $\pm$ 0.28
$^{a}$ \\ \hline
$D^{0} - D^{+}$   & $\Delta^{0}_{c}$  &           -4.53 & -4.77 $\pm$ 0.27 \\
$D^{*0} - D^{*+}$ & $\Delta^{1}_{c}$  &           -1.60 & -2.9 $\pm$ 1.3 \\
$D^{*0} - D^{0}$  & $\Delta^{*}_{cu}$ &           143.0 & 142.5 $\pm$ 1.3 \\
$D^{*+} - D^{+}$  & $\Delta^{*}_{cd}$ &           140.1 & 140.6 $\pm$ 1.9
$^{a}$ \\ \hline
$B^{+} - B^{0}$   & $\Delta^{0}_{b}$  &           -0.89 & -0.1 $\pm$ 0.8 \\
$B^{*+} - B^{*0}$ & $\Delta^{1}_{b}$  &            0.02 & NA \\
$B^{*+} - B^{+}$  & $\Delta^{*}_{bu}$ &            46.5 & 46.0 $\pm$ 0.6 $^{b}$
\\
$B^{*0} - B^{0}$  & $\Delta^{*}_{bd}$ &            45.5 & 46.0 $\pm$ 0.6 $^{b}$
\\ \hline
\end{tabular}
\\ $^{a}$ obtained as a difference of world averages
\\ $^{b}$ average of charged and neutral states
\end{table}

\clearpage
\section*{Figure Captions}
FIG. 1.  Diagrammatical representation of ${\cal M}_{fi}$. \\ [3ex]
FIG. 2.  Diagram for ${\cal M}_{fi}$ in the mesonic system.\\ [3ex]
FIG. 3.  Free quark Feynman amplitude $\cal M$.\\ [3ex]
FIG. 4.  Notation and conventions for the mesonic system.

\setlength{\unitlength}{5pt}
\clearpage
\begin{picture}(48,36)
\multiput(0,25)(0,1){3}{\line(1,0){6}}
\put(8,26){\oval(4,20)}
\multiput(8,16)(0,20){2}{\line(1,0){10}}
\multiput(10,20)(0,4){4}{\line(1,0){8}}
\put(18,16){\dashbox{2}(12,20){interaction}}
\multiput(30,16)(0,20){2}{\line(1,0){10}}
\multiput(30,20)(0,4){4}{\line(1,0){8}}
\put(40,26){\oval(4,20)}
\multiput(42,25)(0,1){3}{\line(1,0){6}}
\put(2,10){\vector(0,1){14}}
\put(8,13){\vector(0,1){2}}
\put(2,8){composite system}
\put(8,11){superposition}
\put(34,14){\vector(0,1){9}}
\put(36,14){\vector(0,1){5}}
\put(30,12){constituents}
\put(19,4){\bf Fig. 1}
\end{picture}
\vspace{90pt}

\begin{picture}(40,36)
\multiput(0,25)(0,1){3}{\line(1,0){6}}
\put(8,26){\oval(4,20)}
\multiput(8,16)(0,20){2}{\vector(1,0){14}}
\multiput(22,16)(0,20){2}{\line(1,0){10}}
\multiput(20,18)(0,8){3}{\oval(4,4)[l]}
\multiput(20,22)(0,8){2}{\oval(4,4)[r]}
\put(32,26){\oval(4,20)}
\multiput(34,25)(0,1){3}{\line(1,0){6}}
\put(22,26){g,$\gamma$}
\put(14,33){$\bar{Q}$}
\put(14,17){$q$}
\put(2,14){\vector(0,1){10}}
\put(2,12){meson}
\put(32,13){\vector(0,1){2}}
\put(20,11){meson wavefunction}
\put(16,4){\bf Fig. 2}
\end{picture}
\hspace{30pt}
\begin{picture}(24,36)
\multiput(0,16)(0,20){2}{\vector(1,0){14}}
\multiput(14,16)(0,20){2}{\line(1,0){10}}
\multiput(12,18)(0,8){3}{\oval(4,4)[l]}
\multiput(12,22)(0,8){2}{\oval(4,4)[r]}
\put(16,26){g,$\gamma$}
\put(4,33){$\bar{Q}$}
\put(4,17){$q$}
\put(7,6){\bf Fig. 3}
\end{picture}

\clearpage
\begin{picture}(76,36)
\multiput(0,25)(0,1){3}{\line(1,0){6}}
\put(8,26){\oval(4,20)}
\multiput(8,16)(0,20){2}{\vector(1,0){14}}
\multiput(22,16)(0,20){2}{\line(1,0){10}}
\multiput(20,18)(0,8){3}{\oval(4,4)[l]}
\multiput(20,22)(0,8){2}{\oval(4,4)[r]}
\put(32,26){\oval(4,20)}
\multiput(34,25)(0,1){3}{\line(1,0){6}}
\put(0,28){${\cal P} \rightarrow$}
\put(35,28){${\cal P} \rightarrow$}
\put(22,26){$\uparrow k$}
\put(14,17){$p_{i}$}
\put(14,33){$P_{i}$}
\put(26,17){$p_{f}$}
\put(26,33){$P_{f}$}
\put(35,16){$q$, helicity $h$}
\put(35,34){$\bar{Q}$, helicity $H$}
\put(51,31){$\vec{p}_{i} \, = \frac{\vec{\cal P}}{2} + \vec{p}$}
\put(51,27){$\vec{P}_{i}    = \frac{\vec{\cal P}}{2} - \vec{p}$}
\put(51,23){$\vec{p}_{f} \, = \frac{\vec{\cal P}}{2} + \vec{p} \, '$}
\put(51,19){$\vec{P}_{f}    = \frac{\vec{\cal P}}{2} - \vec{p} \, '$}
\put(67,31){$p_{i}^{0} \, \equiv \e _{i}$}
\put(67,27){$P_{i}^{0}    \equiv E_{i}$}
\put(67,23){$p_{f}^{0} \, \equiv \e _{f}$}
\put(67,19){$P_{f}^{0}    \equiv E_{f}$}
\put(58,15){$p_{i}^{2} \, = \, p_{f}^{2} \equiv m^{2}$}
\put(58,11){$P_{i}^{2}    =    P_{f}^{2} \equiv M^{2}$}
\put(18,6){\bf Fig. 4}
\end{picture}

\begin{thebibliography}{99}
\bibitem{Singh}C. P. Singh, A. Sharma, and M. P. Khanna, Phys. Rev. {\bf D 24},
788 (1981).
\bibitem{Chan}L.-H. Chan, Phys. Rev. Lett. {\bf 51}, 253 (1983).
\bibitem{Kim}D. Y. Kim and S. N. Sinha, Ann. Phys. {\bf 42}, 47 (1985).
\bibitem{Flamm}D. Flamm, F. Sch\"{o}berl, and H. Uematsu, Nuovo Cimento
{\bf 98A}, 559 (1987).
\bibitem{DeRujula}A. De R\'{u}jula, H. Georgi, and S. L. Glashow, Phys. Rev.
{\bf D 12}, 147 (1975).
\bibitem{Tiwari}K. P. Tiwari, C. P. Singh, and M. P. Khanna, Phys. Rev.
{\bf D 31}, 642 (1985).
\bibitem{Goity}J. L. Goity and W.-S. Hou, Phys. Lett. {\bf 282B}, 243 (1992).
\bibitem{Bethe}H. A. Bethe and E. E. Salpeter, {\em Quantum Mechanics of
One- and Two-Electron Atoms} (Plenum, New York, 1977).
\bibitem{Gupta}S. N. Gupta, {\em Quantum Electrodynamics} (Gordon and Breach,
New York, 1977), pp. 191-206.
\bibitem{Jacob}M. Jacob and G. C. Wick, Ann. Phys. {\bf 7}, 404 (1959).
\bibitem{Quigg}C. Quigg and J. L. Rosner, Phys. Rep. {\bf 56C}, 167 (1979).
\bibitem{Eichten}E. Eichten, K. Gottfried, T. Kinoshita, K. D. Lane, and T.-M.
Yan, Phys. Rev. {\bf D 17}, 3090 (1978).
\bibitem{Olsson}M. G. Olsson, University of Wisconsin, Madison Report No.
MAD/PH/656, June, 1991 (unpublished).
\end{thebibliography}
\end{document}